\documentclass[aps,preprint,showpacs,preprintnumbers,amsmath,amssymb]{revtex4}
\usepackage{graphicx}
\begin{document} 
\def\beq{\begin{equation}}
\def\eeq{\end{equation}}
\def\beqn{\begin{eqnarray}}
\def\eeqn{\end{eqnarray}}

\title{Build-up of coherence between initially-independent subsystems: The case of Bose-Einstein condensates}

\author{O. E. Alon, A. I. Streltsov, and L. S. Cederbaum}
\affiliation{Theoretische Chemie, Physikalisch-Chemisches Institut, Universit\"at Heidelberg,\\
Im Neuenheimer Feld 229, D-69120 Heidelberg, Germany}

\begin{abstract}
When initially-independent subsystems are made to contact,
{\it coherence} can develop due to interaction between them.
We exemplify and demonstrate this paradigm through several scenarios of 
two initially-independent Bose-Einstein condensates which 
are allowed to collide.
The build-up of coherence depends strongly on time, 
interaction strength and other parameters of each condensate.
Implications are discussed.
\end{abstract}
\maketitle

Coherence is a fundamental property in quantum mechanics,
and is the basis for both fundamental and practical research fields in modern physics \cite{C1,C2}.
Often one speaks of the {\it loss} of coherence when a system is coupled to an environment, or a bath. 
In this context and for physical applications relying on coherence, 
one is also interested in how to {\it protect} coherence of quantum systems \cite{C3}.

The issue we address in the present work is a complementary paradigm, namely {\it build-up} of coherence.
We show that when initially-independent subsystems are made to contact,
mutual coherence can develop due to interaction between them.
We exemplify and demonstrate this paradigm through several scenarios of 
two initially-independent Bose-Einstein condensates (BECs) which 
are allowed to collide.
It has been suspected before that coherence builds-up in the collision between two
initially-independent BECs \cite{CH1}.
Here, for the first time,
we show explicitly by solving the time-dependent many-body Schr\"odinger equation 
and resorting to the reduced one-particle density matrix that coherence can build-up and how it happens.
The build-up of coherence depends strongly on time, 
interaction strength and other parameters of each BEC.

We consider two initially-independent BECs, an $A$ BEC with $N_A$ atoms
and a $B$ BEC with $N_B$ atoms. 
In the simplest case,
each of the two BECs is a weakly interacting Bose gas
associated with an orbital,
$\phi_A(x)$ for the $A$ BEC and $\phi_B(x)$ for the $B$ BEC. 
Being independent means that the two BECs do not overlap in space,
$\int dx|\phi_A(x)|^2|\phi_B(x)|^2=0$.
Nowadays, independent BECs can readily be realized if the two BECs are held in two different traps,
or in a double-well potential with a large barrier between the wells.
When the $A$ and $B$ BECs are comprised of the same kind of bosons, $N=N_A+N_B$,
the quantum state of the whole system reads $\Psi=\hat{\cal S}\phi_A(x_1)\ldots\phi_A(x_{N_A}) 
\phi_B(x_{N_A+1})\ldots\phi_B(x_{N_A+N_B})$, 
where $\hat{\cal S}$ is the symmetrization operator.
The many-boson state $\Psi$ thus describes a fragmented BEC \cite{N1,S1,AO1,E1}.

At time $t=0$ we remove the traps and allow the whole system to evolve in free space
under the many-body Hamiltonian $\hat H = \sum_j \hat T(x_j) + \sum_{j<k} \hat W(x_j-x_k)$.
Here, $\hat T(x)$ is the kinetic energy operator and $\hat W(x-x')$ the inter-particle interaction.
The system is taken to be one-dimensional. 
The many-body state of the system evolves according to the time-dependent Schr\"odinger equation
$\hat H\Psi(t)=i\frac{\partial\Psi(t)}{\partial t}$.
We solve the Schr\"odinger equation with the recently developed
multi-configurational time-dependent Hartree for bosons (MCTDHB) \cite{M1}.
In the MCTDHB($M$) the time-dependent many-boson wavefunction $\Psi(t)$ is written
as a linear combination of all possible permanents resulting
by distributing $N$ bosons over $M$ time-dependent orthogonal orbitals.
The MCTDHB($M$) wavefunction reads
$\Psi(x_1,x_2,\ldots,x_N,t) = \sum_{\vec{m}} C_{\vec{m}}(t)
\hat{\cal S}\phi_1(x_1,{{t}})\cdots\phi_2(x_{m_1+1},{t})
\cdots\phi_3(x_{m_1+m_2+1},{t})
\cdots\phi_M(x_{N},{t})$,
where $\vec{m}=(m_1,m_2,\cdots m_M)$ collects the occupation numbers.
Within this theory a quantitative description of the time evolution of bosonic systems is achieved by optimizing
all $M$ orbitals used to construct the many-body expansion and the expansion coefficients themselves
at each point in time utilizing a standard time-dependent variational principle \cite{M1}.
To analyze the system's evolution we will also resort to the respective time-dependent 
multi-orbital mean-field (TDMF) dynamics \cite{T1} of the two initially-independent BECs \cite{I0,I1}.
In TDMF($M$) theory the many-boson wavefunction has the form
$\Psi(x_1,\ldots,x_N,t) = \hat{\cal S}\phi_1(x_1,{{t}})\cdots\phi_2(x_{m_1+1},{t})
\cdots\phi_3(x_{m_1+m_2+1},{t}) \cdots\phi_M(x_{N},{t})$, i.e.,
the time-dependent many-body wavefunction is comprised
of one permanent only and thus maintains the {\it initial} occupation numbers $\vec{m}$ in time.
Obviously, the many-body dynamics goes much beyond the TDMF one.

For a system comprised of identical bosons the most
basic quantity quantifying coherence in the system is
the reduced one-body density matrix 
$\rho(x,x',t)= N \int dx_2 \ldots \int dx_N |\Psi^\ast(x',x_2,\ldots,x_N,t)\Psi(x,x_2,\ldots,x_N,t)|$.
The reduced one-body density matrix has the familiar spectral resolution 
$\rho(x,x';t) = \sum_j n_j(t) \psi^\ast_j(x',t)\psi_j(x,t)$
where $n_j(t)$ are the natural occupation numbers and $\psi_j(x,t)$ are the natural orbitals.
For the scenarios studied here there are two principal natural orbitals occupied.
Furthermore, we will discuss the density $\rho(x,t)=\rho(x,x=x',t)$,
namely the diagonal part of the reduced one-body density matrix.

We consider two-initially independent BECs with $N_A=N_B=500$ bosons each.
This system is prepared as the ground-state 
of a symmetric double-well potential formed by bisecting an harmonic potential with a Gaussian-shaped barrier.
Here and hereafter we work in dimensionless units which are
arrived at by choosing a convenient length scale $L$ (say the initial distance between the BECs) 
and dividing the Hamiltonian $\hat H$ by $\frac{\hbar^2}{m L^2}$ where $m$ is the boson mass. 
The kinetic energy operator then reads $\hat T = -\frac{1}{2}\frac{\partial^2}{\partial x^2}$,
and the double-well is $\hat V(x)=0.05x^2+50/\sqrt{2\pi}\exp[-x^2/8]$.
As the particle-particle interaction we employ the standard delta-function potential,
$\hat W(x-x')=\lambda_0\delta(x-x')$, where the transverse 
confinement is accounted for in $\lambda_0$ \cite{O1}. 
Here $\lambda_0=0.1$ 

\begin{figure}[ht] 
\includegraphics[width=11cm,angle=-90]{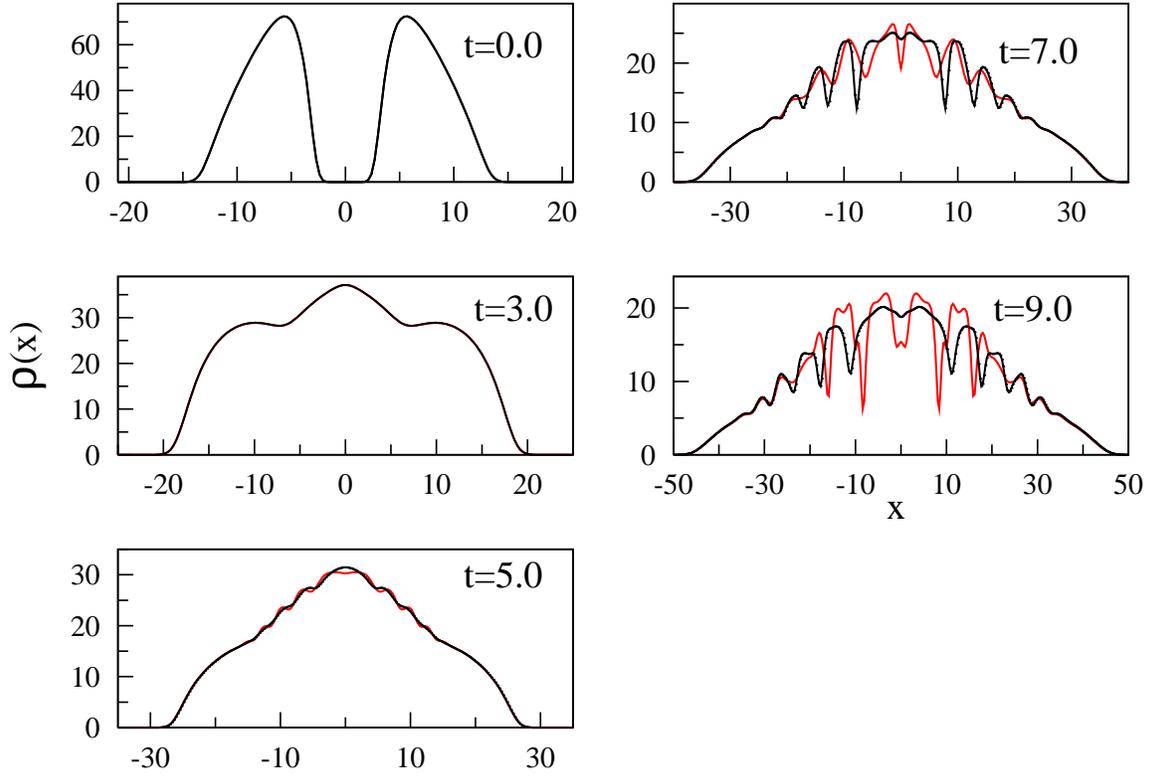}
\caption [kdv]{(Color online) Free expansion and interferences of two-initially independent BECs
each made of $500$ bosons for $\lambda_0=0.1$.
Shown as a function of time are snapshots of the density.
On the multi-orbital mean-field level [$\rho^{MF}(x,t)$ -- red curves] interferences 
result from ``interaction-assisted self-interference'' mechanism of the two BECs \cite{I1}.
On the many-body level [$\rho(x,t)$ -- black curves] there is in addition build-up
of coherence in the system. 
See text for more details.
The quantities shown are dimensionless.
}
\label{F1}
\end{figure}

In Fig.~\ref{F1} we present snapshots of the many-body density $\rho(x,t)$.
Also shown is the mean-field density $\rho^{MF}(x,t)$ computed by the TDMF approach.
Both densities show interferences of the two initially-independent BECs,
a subject which has drawn much attention recently \cite{CH1,I0,I1,B1}.
At short times, up to about $t=3.0$, the two densities and thus the interference patterns coincide.
Thereafter, we see that differences start to appear between the two densities.
It has been shown recently that interaction leads to interferences
in the density of two initially-independent BECs \cite{CH1,I0,I1}.
The interaction between the two expanding BECs leads to 
``interaction-assisted self-interference'' mechanism between them \cite{I1}.
This mechanism is seen in the mean-field density $\rho^{MF}(x,t)$ of Fig.~\ref{F1}.
The differences between the many-body and mean-field densities
seen in Fig.~\ref{F1} thus indicate that
more than ``interaction-assisted self-interference'' is happening in the system.

\begin{figure}[ht] 
\includegraphics[width=11cm,angle=-90]{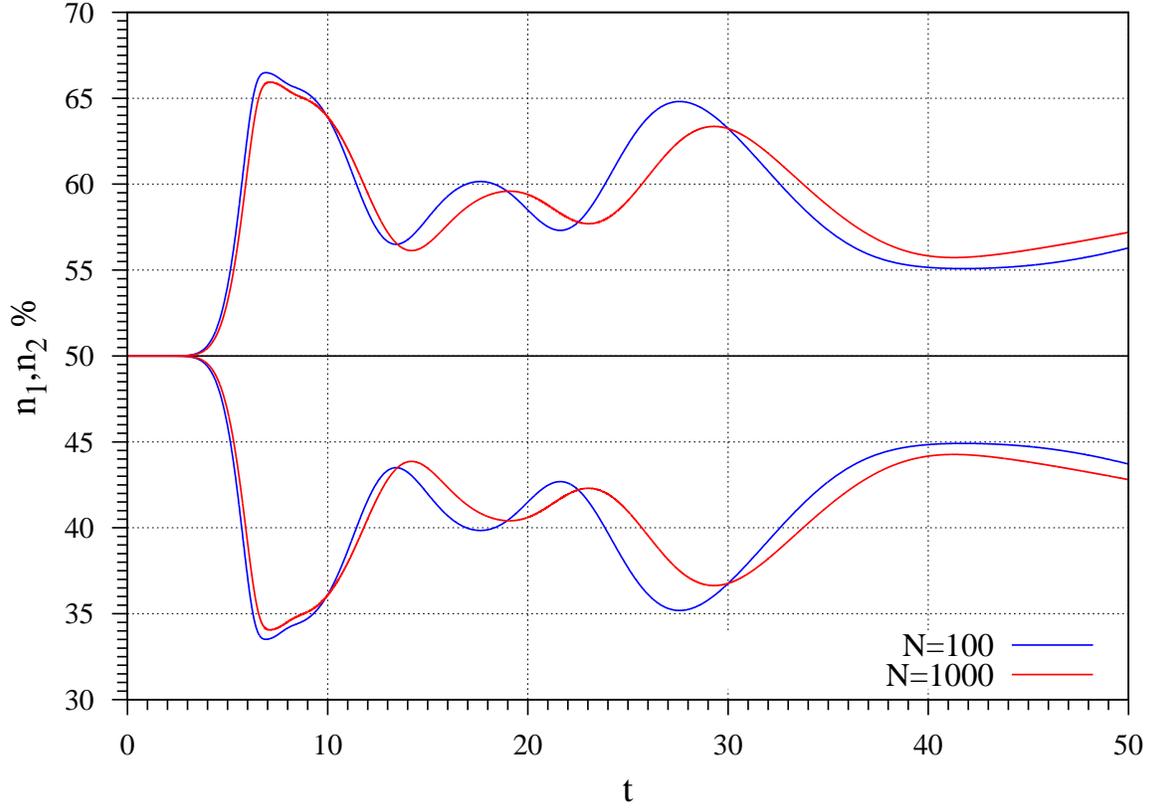}
\caption [kdv]{(Color online) 
Build-up of coherence between two initially-independent BECs.
Shown are the natural occupation numbers $n_j(t)$ as a function of time
for the system in Fig.~\ref{F1}.
For comparison, the evolution of $n_j(t)$ for two initially-independent BECs 
with $50$ bosons each and the same $\lambda_0(N-1)$ factor is shown. 
The quantities shown are dimensionless.}
\label{F2}
\end{figure}

To understand and quantify the differences between the many-body and mean-field 
densities we begin by plotting in Fig.~\ref{F2} the natural occupation numbers $n_j(t)$ of the many-body solution. 
We recall that in the mean-field dynamics the natural occupation numbers do not change in time. 
The natural occupations numbers $n_j(t)$ stay ``flat'' at their initial values up to about $t=4$. 
Note that the two initially-independent BECs have already met at that time, see Fig.~\ref{F1}.
From about $t=4$ on,
the natural occupation numbers change from their initial values and {\it coherence} develops in the system.
We see that coherence develops in an oscillatory manner and not monotonously.
At about $t=7$ the coherence assumes its maximal value
for the system under investigation.
It is instructive to employ the quantity 
$C(t)=100 \times \frac{n_1(t)-n_1(0)}{N-n_1(0)}$ which indicates the {\it fraction} in percents
at time $t$ of the {\it maximally} possible coherence in the system.
By definition, $C=0$ at $t=0$.
In our system of two initially-dependent BECs with the same number of bosons 
$C(t)$ takes a simplified and appealing form, $C(t)=100 \times \frac{n_1(t)-n_2(t)}{n_1(t)+n_2(t)}$, 
measuring the {\it total} coherence in the system.
For two initially-independent BECs $C=0$,
and if the whole system becomes condensed $C=100\%$.
From Fig.~\ref{F2} we see that in our example the maximal coherence 
developed ($t\approx 7$) in the whole system is about $C=34\%$.

The result depicted in Fig.~\ref{F2} clearly shows that coherence builds-up in the system
as a consequence of the interaction between the initially-independent BECs.
The coherence changes with time in a nontrivial oscillatory manner.
We have also examined the build-up of coherence in a smaller system of $N=100$ bosons.
We considered two initially-independent BECs with $N_A=N_B=50$ bosons prepared in the ground-state
of the above double-well potential $V(x)$.
The interaction strength $\lambda_0$ is tuned such that
the factor $\lambda_0(N-1)$ is the same as for the above-studied
case with $N=1000$ atoms. 
The build-up of coherence is clearly
observed in this smaller system as seen in Fig.~\ref{F2}. 
It is interesting to note
that the build-up of coherence of the two systems 
follows a similar time-dependent pattern.

It is instructive to return to the density in Fig.~\ref{F1}
and analyze how the build-up of coherence influences the density $\rho(x,t)$.
As there are two principal natural occupation numbers the density is expressed as
\beq\label{e1}
 \rho(x,t) = n_1(t) |\psi_1(x,t)|^2 + n_2(t) |\psi_2(x,t)|^2.
\eeq
The initial conditions of the two initially-independent BECs are: 
$n_1(0)=N_A, \psi_1(x,0)=\phi_A(x)$ and $n_2(0)=N_B, \psi_2(x,0)=\phi_B(x)$.
With time, the interaction between the bosons
modifies the natural orbitals $\psi_1(x,t), \psi_2(x,t)$
as well as the natural occupation numbers $n_1(t), n_2(t)$.
Similarly, within the TDMF approach \cite{T1}
the density is composed of two natural orbitals as follows
\beq\label{e2}
 \rho^{MF}(x,t) = N_A |\phi_A(x,t)|^2 + N_B |\phi_B(x,t)|^2.
\eeq
The initial conditions are the same,
$\phi_A(x,t)=\phi_A(x)$ and $\psi_2(x,0)=\phi_B(x)$,
but the occupation numbers {\it remain} fixed-in-time.

To show how many-body physics and the build-up of coherence
manifest themselves in the time-dependent density,
it is deductive to relate the two densities $\rho(x,t)$ and $\rho^{MF}(x,t)$.
For this,
we expand the many-body natural orbitals $\psi_1(x,t), \psi_2(x,t)$
which are, of course, normalized and orthogonal to one another 
with the help of the TDMF orbitals $\phi_A(x,t), \phi_B(x,t)$.
Specifically, we write
$\psi_1(x,t)=\sqrt{1-|\Delta_{1B}(t)^2|-|\Delta_{1Q}(t)|^2}\phi_A(x,t)+\Delta_{1B}(t)\phi_B(x,t)+\Delta_{1Q}(t)\phi_{1Q}(x,t)$
and
$\psi_2(x,t)=\sqrt{1-|\Delta_{2A}(t)|^2-|\Delta_{2Q}(t)|^2}\phi_B(x,t)+\Delta_{2A}(t)\phi_A(x,t)+\Delta_{2Q}(t)\phi_{2Q}(x,t)$.
The first two terms in each expansion are the TDMF orbitals which are themselves orthogonal to one another. 
The remaining term of each expansion $\phi_{1Q}(x,t), \phi_{2Q}(x,t)$
belongs to the subspace orthogonal to TDMF orbitals.
The functions $\phi_{1Q}(x,t)$ and $\phi_{2Q}(x,t)$ are in general non-orthogonal.
Similarly, we write for the time-dependent natural occupations 
$n_1(t) = N_A + \Delta n(t)$ and $n_2(t) = N_B - \Delta n(t)$.
Thus, by construction, at $t=0$ all quantities denoted with $\Delta$ 
are equal to zero; 
Their non-zero values obtained with time signify {\it many-body} 
facets of the interaction in the system.  

It is straightforward to insert the above expansions for the natural orbitals and occupation numbers into Eq.~(\ref{e1}). 
The expression obtained is lengthy and at first sight not very informative.
More informative is to analyze $\rho(x,t)$ to first order in the quantities denoted in the expansions by $\Delta$.
The result takes on the following form:
\beqn\label{e3}
 \rho(x,t) &=& \rho^{MF}(x,t) + \Delta n(t)\left[|\phi_A(x,t)|^2-|\phi_B(x,t)|^2\right] \nonumber \\
&+& 2{\mathrm Re}\left\{N_A\left[\Delta_{1B}\phi^\ast_A(x,t)\phi_B(x,t)\right] + 
  N_B\left[\Delta_{2A}\phi^\ast_B(x,t)\phi_A(x,t)\right] \right\} \nonumber \\
&+& 2{\mathrm Re}\left\{N_A\left[\Delta_{1Q}\phi^\ast_A(x,t)\phi_{1Q}(x,t)\right] + 
  N_B\left[\Delta_{2Q}\phi^\ast_B(x,t)\phi_{2Q}(x,t)\right]\right\}. \
\eeqn
With its four terms, 
Eq.~(\ref{e3}) offers an appealing and explicit {\it mechanism} for
the many-body effects of interaction on the density of two initially-independent BECs.
The first term is just the mean-field density $\rho^{MF}(x,t)$.
As discussed above and shown in Fig.~\ref{F1},
at short times this is the only (visible) contribution to the density.
As time progresses corrections to the mean-field quantities start to mount, 
also see Fig.~\ref{F1},
resulting in three additional contributions to $\rho(x,t)$.
The second term in (\ref{e3}) is due to the build-up of coherence in the system,
and describes the flow of particles from one BEC to another without changing the TDMF orbitals themselves.
The third term in (\ref{e3}) is an {\it interference} term
and describes interference between the two BECs $\phi_A(x,t)$ and $\phi_B(x,t)$.
It originates from the change of the many-body natural orbitals {\it within} 
the TDMF subspace $\{\phi_A(x,t), \phi_B(x,t)\}$, 
without changing the TDMF occupation numbers.
The last, fourth term is another interference term.
The many-body natural orbitals start to change form
{\it beyond} the linear-combinations of TDMF orbitals.
The fourth term thus describes the interference of the $A$ BEC [$\phi_A(x,t)$]
with $\phi_{1Q}(x,t)$ and, separately,
the interference of the $B$ BEC [$\phi_B(x,t)$] with $\phi_{2Q}(x,t)$.
Adding up the above four contributions, 
the time-dependent density of two initially-independent BECs
reflects combined interference and build-up-of-coherence mechanisms.
Finally, we remark that going beyond first order in the $\Delta$ terms,
see Eq.~(\ref{e3}), there are more interference terms and 
terms resulting from the build-up of coherence.
Of course, the many-body dynamics computed with the MCTDHB approach 
and depicted in the figures are to all orders in $\Delta$.
The above analysis in $\Delta$ is for the sake of physical interpretation only.

\begin{figure}[ht] 
\includegraphics[width=11cm,angle=-90]{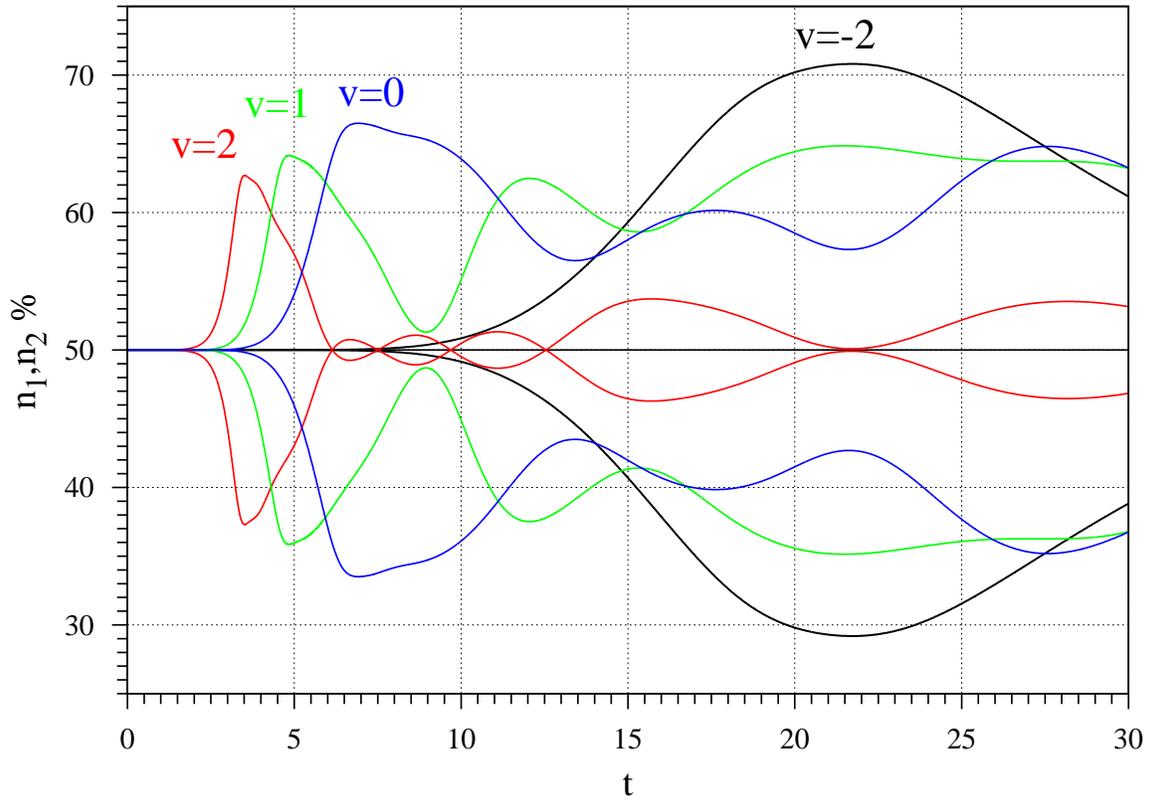}
\caption [kdv]{Influence of relative velocity $2 \times v$ on the build-up of coherence.
Positive $v$ indicates that the BECs move towards each other and negative $v$ the reverse.
Shown are the natural occupation numbers $n_j(t)$ as a function of time
for two initially-independent BECs with $50$ bosons each. 
The quantities shown are dimensionless.}
\label{F3}
\end{figure}

Next, we would like to study more on the development of coherence in a system
with two initially-independent BECs.
We have seen that coherence develops in time and the question we would
like to address is whether we can influence this process.
We make the following 'experiment'.
We release the two BECs and give the $A$ BEC velocity $v$ and the $B$ BEC the opposite velocity $-v$.
Positive $v$ means that the BECs move towards each other and negative $v$ the reverse.
The relative velocity of the two BECs is of course $2\times v$. 
Examples are collected in Fig.~\ref{F3} where the 
natural occupation numbers $n_j(t)$ of the many-body solution are plotted.
The faster the two BECs move towards each other,
the quicker coherence initially develops 
(see the maximal $n_1(t)$ at about $t=7,5,3$ for velocities $v=0,1,2$, respectively).
This behavior can be anticipated by the kinematics:
the faster the BECs move towards each other,
the sooner they come in contact which allows coherence to build-up.
Side-by-side, we observe that the coherence oscillates and with increasing positive velocity $v$,
the height of the first oscillation decreases.
An interesting point seen in Fig.~\ref{F3} for positive $v$ is that 
this decrease is approximately linear in $v$.
Let us make another 'experiment', in which the two initially-independent BECs move {\it away} from one another.
The natural occupation numbers are plotted for $v=-2$ in Fig.~\ref{F3}.
We first see that coherence develops slower than for positive $v$, 
in conjunction with the above kinematic analysis. 
However, its maximum ($C\approx 42\%$) is even higher than the maximal amount of coherence developed with
initially-independent BECs at rest ($v=0$).
To explain this, at first sight counterintuitive result, 
we recall that the build-up of coherence between initially-independent BECs
implies that more bosons share the same natural orbital $\psi_1(x,t)$.
How to 'help' initially-independent bosons share the same natural orbital?
Bosons coming from initially-independent BECs at rest
have positive relative velocity due the free expansion of the two BECs.
These bosons are more `distinct' from one another and therefore less coherence can build-up.
Sending the two BECs with negative relative velocity 
compensates partly for this expansion of the two BECs.
The now slower bosons are more 'similar' to one another and 
therefore higher degree of coherence builds-up.

Let us summarize.
Coherence can develop between two initially-independent BECs.
How it develops in time when the two BECs are released 
from their trap strongly depends on different parameters,
such as the inter-particle interaction and relative velocity between the BECs.
We have considered the build-up of coherence
between two initially-independent BECs comprised of indistinguishable particles.
This motivates the consideration of other systems made of initially-independent subsystems 
where we also may anticipate build-up of coherence in time.
For instance, the possible build-up of coherence between two subsystems,
one coherent the other not, is an interesting case.
In the context of quantum gases,
the two subsystems can be a BEC and 
a Mott-insulator in an optical lattice \cite{P1,IB1}.
Another appealing direction is when the initially-independent
subsystems are made of different kinds of particles.
Here, due to the distinguishablity of the particles,
it would be instructive to enquire on the possible build-up of coherence 
within each subsystem as well as in the system as a whole.
The latter is measured by higher-order density matrices.

\acknowledgments
\noindent
Financial support by DFG is acknowledged.

\end{document}